# Concepts and Applications of Conformal Prediction in Computational Drug Discovery

Isidro Cortés-Ciriano[1,*] and Andreas Bender[2]

[1] European Molecular Biology Laboratory, European Bioinformatics Institute, Wellcome Genome Campus, Hinxton CB10 1SD, UK

[2] Centre for Molecular Informatics, Department of Chemistry, University of Cambridge, Lensfield Road, Cambridge, CB2 1EW, United Kingdom.

[*]Corresponding author: icortes@ebi.ac.uk

**Abstract**

Estimating the reliability of individual predictions is key to increase the adoption of computational models and 'artificial intelligence' in preclinical drug discovery, as well as to foster its application to guide decision making in clinical settings. Among the large number of algorithms developed over the last decades to compute prediction errors, Conformal Prediction (CP) has gained increasing attention in the computational drug discovery community. A major reason for its recent popularity is the ease of interpretation of the computed prediction errors in both classification and regression tasks. For instance, at a confidence level of 90% the true value will be within the predicted confidence intervals in at least 90% of the cases. This so called *validity* of conformal predictors is guaranteed by the robust mathematical foundation underlying CP. The versatility of CP relies on its minimal computational footprint, as it can be easily coupled to any machine learning algorithm at little computational cost. In this review, we summarize underlying concepts and practical applications of CP with a particular focus on virtual screening and activity modelling, and list open source implementations of relevant software. Finally, we describe the current limitations in the field, and provide a perspective on future opportunities for CP in preclinical and clinical drug discovery.

## Introduction

A major research area in machine learning is the development of algorithms to compute the reliability of individual predictions. Such reliability estimates are essential to increase the trust and application of artificial intelligence solutions to guide decision making, especially in the context of healthcare and personalized medicine, where correctly identifying *e.g.,* which patients are likely to benefit from a particular drug treatment has strong ethical and legal implications[1]. Estimating the reliability of predictive models is of particular relevance in drug discovery, where both the prediction and the associated uncertainty need to be taken into account for decision making[2]. The set of molecules for which a model is expected to generate reliable predictions is termed the *applicability domain* of a model[3]. Therefore, the development of algorithms to define and compute the applicability domain of predictive models has been an area of intense research in preclinical drug discovery, and due to the amount of recent research in the field we will in the following often focus on virtual screening in particular[4–7].

The generation of a predictive model, such as a Quantitative-Structure Activity Relationship (QSAR) model, or any other property model, consists of encoding the set of molecules for which the variable of interest has been experimentally measured (*e.g., in vitro* potency; the dependent or response variable) using numerical descriptors (covariates or independent variables)[8–10]. Subsequently, the covariates are related to the response variable using a mathematical model, which can be simple (*e.g*., multiple linear regression) or complex, such as Random Forests (RF), Support Vector Machines (SVM) or Deep Neural Networks, which might consist of thousands to millions of parameters[11]. These models are then used to predict the activity or property of interest for untested molecules *in silico* to prioritize for further experimental testing those with higher chances of being active. Although a plethora of algorithmically diverse approaches to encode chemical structures and to relate these to a dependent variable of interest have been developed over the last decades[10,12], all depend on the limited amount of data available for training. This is key to drug discovery, as the amount and diversity of the training data (in this case chemical compounds) limits the set of small molecules to which a given model can be applied and give reasonable results. In practice usually more (and more homogenous) data is available in early-stage settings (such as hit discovery), and less so in later (such as *in vivo*) stages, making early stage model generation generally speaking relatively easier than late-stage models. However, what is generally applicable and a key question is how to assign the confidence, reliability, or applicability domain of a model, and while quantitatively models in different stages might have larger or smaller applicability domains its importance is universal.

Widely-used algorithms to generate QSAR models, such as Support Vector Machines (SVM) or Random Forest (RF), output a single value (regression) or label (classification) for new instances[13–15]. In practice, the lack of uncertainty estimates for point or single-class predictions hampers the use of computational models to guide prospective screening experiments in early-stage drug discovery, given that 'desired' values of an output variable may well be associated with high uncertainty (and vice versa), but this will not be explicitly communicated to the user by the algorithm. However, in the same way experiments need to be repeated in order to assign both a measurement and its standard deviation of a variable, also computational predictions need to output the expectation value of the model and an associated confidence interval.

We note that there exist algorithms whose outputs are well-calibrated probability distributions rather than point predictions, *e.g.,* Gaussian Processes (GP)[16–19] or Dropout Neural Networks[20,21]. However, these are generally computationally intense (*e.g.,* GP training requires the inversion of the covariance matrix, which drastically increases their computational footprint), require the optimization of a large number of parameters, as well as prior knowledge about these[19]. Hence, most applications in the medicinal chemistry literature use alternative algorithms (*e.g.,* RF), which are faster to train (by *e.g.,* parallelizing the training phase), and require less parameter optimization[22,23]. Therefore, much effort has been invested in the community to develop uncertainty estimation methods that could be easily adapted to algorithms used in practice[22,24–27].

To date, a plethora of diverse algorithms have been developed to define the applicability domain of virtual screening models[3,5,7,10,12,17,18,22,24,25,28–33]. Overall, existing methods harness the distance in descriptor space of new instances to those in the training set [25,34], or intrinsic information derived from the models (*e.g.*, bagged variance, which is positively correlated with high average error in prediction[22,35]), to identify areas in chemical space underrepresented in the training data for which the models are unlikely to deliver reliable predictions. The performance of these approaches is generally quantified by testing whether the error in prediction increases as a function of the distance of the test molecules to those in the training data[34]. However, such correlations do not guarantee that the predictions are well-calibrated, *i.e.,* the fraction of instances whose true value lies within the predicted error interval is not guaranteed to be proportional to a user-defined tolerated error rate.

One method which has recently gained a lot of attention to address this problem is Conformal Prediction (CP; Table 1)[36–38]. CP is a mathematical framework to generate confidence predictors to model the reliability of predictions in diverse tasks, from property modelling to multi-target drug design (Table 2). A confidence predictor is a type of predictor that guarantees that the true value will be within the predicted confidence region (in regression) or within a set of predicted labels (in the case of classification) at a given confidence level (CL). For instance, at a confidence level of 80%, the confidence intervals computed using a valid Conformal Predictor in a regression setting would contain the true value in at least 80% of the cases (Figure 1a). In classification tasks, the set of predicted classes for new instances will contain the true label in at least 80% of the cases. The prediction of well-calibrated (or *valid*) confidence regions by CP, which is usually not the case for other modelling methods, guarantees a lower bound for validation rates, and permits to limit the number of false positives, thus increasing the retrieval rate of active compounds in preclinical drug discovery[39]. The confidence level is commonly set at 0.80, as this confidence level represents a generally suitable trade-off between efficiency and validity (note that a CL of 80% is also reported as 0.80 in the CP literature)[40]. The validity of CP holds if the randomness assumption is fulfilled*, i.e.,* the training instances are representative of those to which the models will be applied; or, in other words, that the data are independent and identically distributed (i.i.d.). In practice, CP are also valid if the slightly weaker assumption of exchangeability is fulfilled, which assumes that the datapoints do not follow any particular order even if they are not i.i.d[36,39]. Although the exchangeability principle is generally assumed to hold when modelling preclinical data[32,36,37], the authors have shown that this is not always the case in virtual iterative screening experiments using QSAR data sets (Figure 1b)[41]. Overall, CP does not introduce more assumptions than those generally made when modelling bioactivity data.

To generate errors associated with predictions, Conformal Predictors evaluate the 'non-conformity' of a new instance to those used during training by applying a new *non-conformity measure*. Intuitively, the non-conformity measure quantifies how different a given instance is from those already seen during the training phase, which is quantified using a *non-conformity score* ($\alpha$). Therefore, any metric quantifying the applicability domain of a model, *e.g.*, the distance of a new instance to the training set, can be used as non-conformity measure. The simplest non-conformity measure in a regression setting would be the unsigned error in prediction computed for *e.g.,* the instances or the hold-out folds in cross-validation. In classification tasks, non-conformity scores are generally computed using a class probability estimation method; for instance, the fraction of trees in a RF model voting for a particular class, the distance to the cut-off value or hyperplane that separates classes in the case of *e.g.,* SVM (Table 2). Note that the selection of a non-conformity measure usually exploits information already provided by the underlying algorithm. For instance, in the case of RF models the fraction of Trees voting for each class in classification, of the bagged variance in regression, are often used as non-conformity measures.

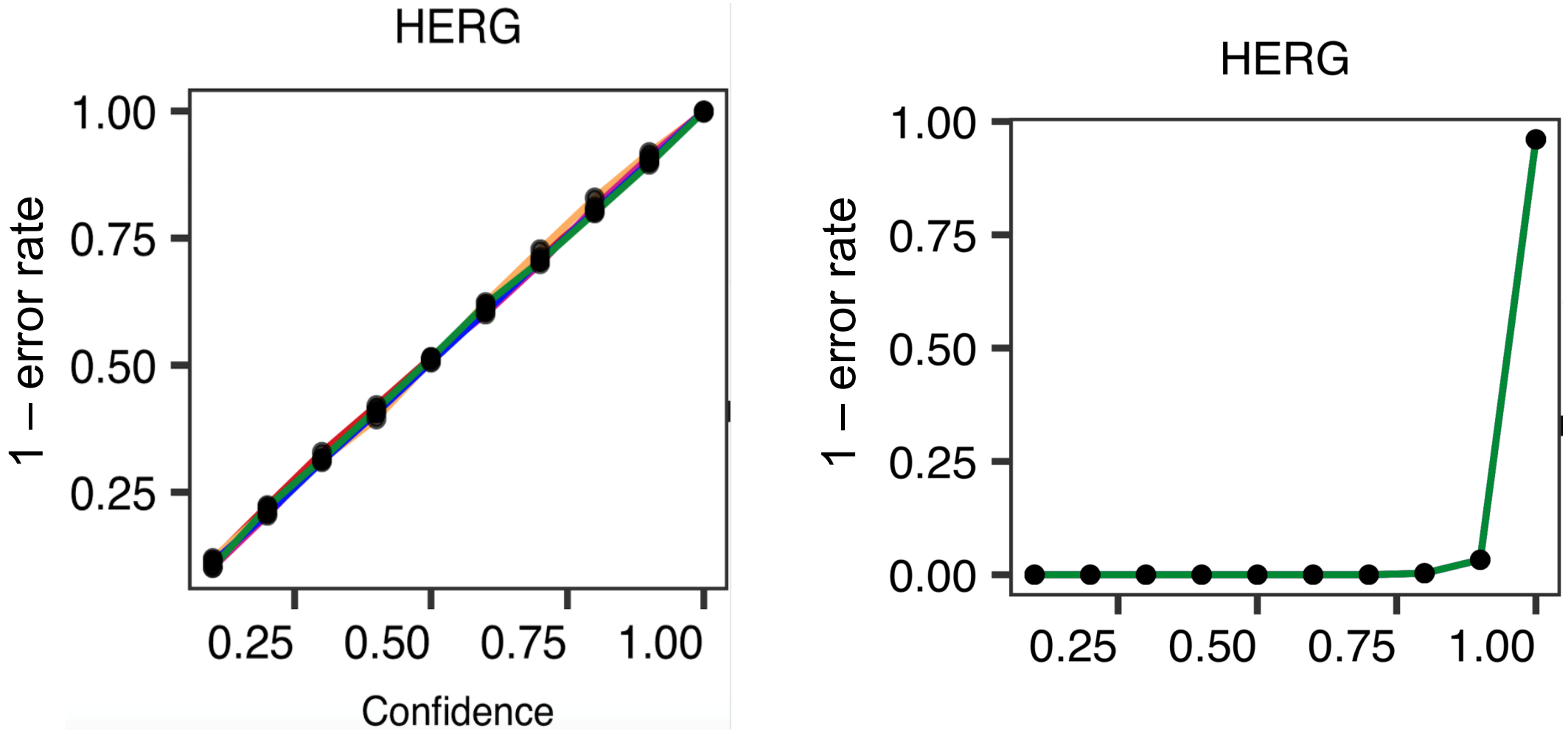

Figure 1. (a) Example of a calibration plot for a well-calibrated Regression Conformal Predictor (adapted from [41]). Calibration plots represent the fraction of confidence regions encompassing the true value (or, 1 – estimated error rate) across increasingly larger confidence levels. The x-axis corresponds to a user-defined confidence level, and the y-axis to the fraction of instances in the test set whose true value lies within the predicted confidence regions. It can be seen that the fraction of instances whose true value is within the predicted confidence interval (y-axis) is correlated with the confidence level, and hence this represents a 'well-calibrated' predictor in the context of Conformal Prediction, which can be used to assign the trust placed in future predictions of the model. (b) Example of a Regression Conformal Predictor that is not well-calibrated (adapted from [41]). It can be seen that the fraction of instances whose true value is within the predicted confidence interval (y-axis) is not correlated with the confidence level.

A second key aspect to consider is the *efficiency* of CP (Table 1). In regression, the efficiency of a Conformal Predictor is determined by the size of the predicted confidence regions. Even if a Conformal Predictor may be valid, it might not be useful in practice to guide decision making if the confidence regions span several bioactivity units[40,42]. In classification, efficiency is related to the fraction of single-class predictions that are correct[38,43].

In addition to the features listed above, Conformal Prediction can be used in combination with any machine learning algorithm[32,37], requires minimal computational cost beyond the training of the underlying algorithm[32], and no parameterization is required except for the selection of a non-conformity measure[40]. Hence, Conformal Prediction is a technique that permits to compute well-calibrated and easy-to-interpret errors in prediction for both balanced and imbalanced data sets, as we discuss more in depth below, at minimal computational cost[44].

The first part of this review provides an overview of the design and advantages of the Conformal Prediction implementations used in drug discovery settings to date. The second part revisits the molecular modelling studies where Conformal Prediction played a key role in either estimating the reliability of individual predictions or in guiding prospective screening experiments. Finally, we discuss the current limitations of Conformal Prediction and offer a perspective on future directions.

Table 1. Common terms used in Conformal Prediction.

| Term | Definition |
| --- | --- |
| Validity | Conformal Predictors are always valid provided that the randomness or exchangeability principles hold. In the case of regression, a Conformal Predictor is valid if the confidence level matches the fraction of instances whose true value lies within the predicted confidence region. For instance, at a confidence level of 80%, the confidence intervals would contain the true value in at least 80% of the cases. In classification tasks, Conformal Predictors are valid in that the set of predicted classes for new instances will contain the true label in at least 80% of the cases. |
| Efficiency | In regression, the efficiency of a CP refers to the average size of the predicted confidence intervals. The tighter the intervals the more efficient a conformal predictor is. In the case of classification, efficiency refers to the fraction of single-class predictions that are correct. |
| Confidence level (CL) | The confidence level is defined by the modeler and refers to the minimum fraction of predictions whose true value will lie within the predicted confidence region, in the case of regression, and the fraction of instances whose true class will be among the set of predicted classes. |
| Error rate | The error rate refers to the fraction of instances whose true value lies outside the predicted confidence regions. If a CP is well-calibrated, the error rate should not be larger than 1-CL (see also Figure 1). |
| Nonconformity measure | Function used to evaluate the relatedness or conformity of new instances to those used for model training. |

Table 2. Drug discovery studies listed in chronological order where Conformal Prediction was implemented.

| Data sets (targets; number of compounds; number of datapoints) | Dependent variable or classes | Machine learning technique | Modelling type and task | Type of Conformal Prediction used | Prospective experimental validation? | Reference | Year | Remarks |
|---|---|---|---|---|---|---|---|---|
| 10 publicly available data sets (Signature descriptors) | $pIC_{50}$ | Random Forest (RF) | QSAR/Regression and Classification | Inductive Mondrian Conformal Prediction | No | Norinder *et al.*[32] | 2014 | Introduction of Conformal Prediction in QSAR modelling |
| 12 human PARPs; 181 compounds; 2,164 datapoints | Thermal shift values measured using differential scanning fluorimetry[45] | RF | PCM/ Regression | Inductive Conformal Prediction | No | Cortés-Ciriano *et al.*[46] | 2015 | Proteochemometric modelling (PCM) of PARP inhibition |
| 2 CAESAR binary classification datasets: carcinogenicity[47] (805 compounds) and mutagenicity[48] (4,204) | Carcinogenic *vs* non-carcinogenic | RF | QSAR/Classification | Inductive Mondrian Conformal Prediction | No | Norinder *et al.*[39] | 2015 | Modelling the carcinogenicity and mutagenicity of compounds using Mondrian Conformal Prediction |
| Ames mutagenicity data set | Non mutagenic *vs* mutagenic | Support Vector Machines (SVM) | QSAR/Classification | Inductive Mondrian Conformal Prediction | No | Ahlberg *et al.*[49] | 2015 | Interpretation of classification models |
| NCI60 data set (59 cancer cell lines; 17,142 compounds; 941,831 datapoints) | $pGI_{50}$ (50% growth inhibition bioassay end-point) | RF | PCM/ Regression | Inductive Conformal Prediction | No | Cortés-Ciriano *et al.*[50] | 2016 | PCM of growth inhibition patterns across the NCI60 data set. Webserver[51] available at www.cclp.marseille.inserm.fr |
| 2 datasets: Estrogen Receptor (binders/non-binders: 445/476) and | Binders vs non-binders (based on $IC_{50}$ and relative binding affinity values) | RF | QSAR/Classification | Aggregated Mondrian Conformal Prediction | No | Norinder *et al.*[54] | 2016 | Modelling compound binding to the estrogen and androgen receptors |

| Data sets (targets; number of compounds; number of datapoints) | Dependent variable or classes | Machine learning technique | Modelling type and task | Type of Conformal Prediction used | Prospective experimental validation? | Reference | Year | Remarks |
|---|---|---|---|---|---|---|---|---|
| Androgen Receptor (binders/ non-binders: 292/633)[52,53] | | | | | | | | |
| The Directory of Useful Decoys, Enhanced (DUD-E)[55] | Actives *vs* inactives | RF | QSAR/Classification | Aggregated Mondrian Conformal Prediction | No | Svensson *et al.*[56] | 2017 | Application of Conformal Prediction to guide docking experiments in an iterative fashion |
| 18 bioactivity data sets from the ExCAPE database[8] | Actives *vs* inactives | SVM | QSAR/ Classification | Mondrian Cross-Conformal Predictors | No | Sun *et al.*[57] | 2017 | Large-scale analysis of the application of Mondrian conformal predictors to model imbalance data sets with imbalance levels of 1:10 to 1:1000. Non-conformity measure based on the SVM decision function |
| 3 datasets from PubChem (AID493091, AID2796 and AID1851) and Ames mutagenicity data set from Hansen *et al.*[58] | Actives *vs* inactives | SVM | QSAR/Classification | Aggregated Mondrian Conformal Prediction | No | Norinder *et al.*[59] | 2017 | Binary classification of imbalanced data using SVM and the distance to the separating hyperplane as the non-conformity measure |
| Data set extracted from Baba *et al.*[60] encompassing 211 compounds | Permeation rate (log Kp) through the human skin | RF and SVM | Regression | Aggregated Mondrian Conformal Prediction | No | Lindh et al.[61] | 2017 | Prediction of the permeation rate (log Kp) of chemical compounds through the human skin |
| 16 cytotoxicity datasets extracted from PubChem (data set size ranging from 29,938 to | Toxic *vs* non-toxic | RF | QSAR/Classification | Aggregated Mondrian Conformal Prediction | No | Svensson *et al.*[62] | 2017 | Large scale modeling of cytotoxicity using largely imbalanced dataset (0.8% average fraction of toxic compounds) |

| Data sets (targets; number of compounds; number of datapoints) | Dependent variable or classes | Machine learning technique | Modelling type and task | Type of Conformal Prediction used | Prospective experimental validation? | Reference | Year | Remarks |
|---|---|---|---|---|---|---|---|---|
| 404,016 compounds) | | | | | | | | |
| 25 protein targets from ChEMBL version 23 | $pIC_{50}$ | RF | QSAR/Regression | Cross-Conformal Prediction | No | Cortés-Ciriano *et al.*[41] | 2018 | Demonstration that the exchangeability principle does not hold when conformal predictors trained on inactive molecules are applied generate CI for active molecules |
| 24 protein targets from ChEMBL version 23 | $pIC_{50}$ | Fully-connected deep neural networks | QSAR/Regression | Deep Confidence | No | Cortés-Ciriano *et al.*[63] | 2018 | Deep Confidence: Conformal Prediction strategy that used snapshot ensembles[64] to generate conformal predictors for deep neural networks |
| 936 primary aromatic amines (630 mutagenic and 306 non-mutagenic) | Mutagenic *vs* non-mutagenic | RF | QSAR/ Classification | Inductive Mondrian Conformal Prediction | No | Norinder *et al.*[65] | 2018 | Prediction of aromatic amine mutagenicity |
| DGM/NIHS data set[66] encompassing 12,140 compounds categorized as: strong mutagenic, mutagenic and non-mutagenic. | Strong mutagenic or mutagenic *vs* non-mutagenic | RF | QSAR/ Classification | Aggregated Mondrian Conformal Prediction | No | Norinder *et al.*[67] | 2018 | Prediction of Ames mutagenicity |
| 316,974 from the PubChem BioAssay database (AID 540275) | Active vs inactive | RF | QSAR. Classification | Aggregated Mondrian Conformal Prediction | No | Norinder *et al.*[68] | 2018 | Virtual screening of vanilloid receptor type 1 (TRPV1) agonists |
| 15,350 data points across 31 | $IC_{50}$, $K_i$, $K_d$, %inhibition and $\Delta Tm$ | RF, SVM and Generalized | PCM. Classification | Mondrian Cross- | Yes | Giblin *et al.*[69] | 2018 | Application of PCM and CP to model the activity of compounds across |

| Data sets (targets; number of compounds; number of datapoints) | Dependent variable or classes | Machine learning technique | Modelling type and task | Type of Conformal Prediction used | Prospective experimental validation? | Reference | Year | Remarks |
|---|---|---|---|---|---|---|---|---|
| bromodomain collected from ChEMBL 20, PubChem, ChEpiMod, GOSTAR, and proprietary AstraZeneca databases | | Linear Models (GLM) | | Conformal Predictors | | | | bromodomains. The model predictions for 1,139 compounds were prospectively validated |
| 29 property, toxicity and bioactivity data sets from ChEMBL version 19 | $pIC_{50}$, logS, Toxicity | RF, Gradient Boosting Machines (GBM), Lasso, Ridge Regression, Bayesian Ridge Regression, Adaptive Boosting (AdaBoost), Automatic Relevance Determination (ARD), Elastic Net, and Partial Least-Squares (PLS) | QSAR and QSPR. Regression | Inductive and Aggregated Conformal Prediction | No | Svensson *et al.*[40] | 2018 | Benchmarking study of non-conformity measures using QSAR data sets |
| 12 PubChem data sets (AIDs: 411, 868, 1030, 1460, 1721, 2314, 2326, 2451, 2551, 485290, 485314, 504444) | Actives *vs* inactives | RF | QSAR/Classification | Aggregated Mondrian Conformal Prediction | No | Svensson *et al.*[70] | 2018 | Application of a gain-cost function including the screening cost and gain in terms of bioactivity information (*e.g.,* discovery of a new hit) |

| Data sets (targets; number of compounds; number of datapoints) | Dependent variable or classes | Machine learning technique | Modelling type and task | Type of Conformal Prediction used | Prospective experimental validation? | Reference | Year | Remarks |
|---|---|---|---|---|---|---|---|---|
| 108 compounds[71] | Skin sensitizers *vs* non-sensitizers | SVM | QSAR/Classification | Aggregated Mondrian Conformal Prediction | No | Forreryd *et al.*[72] | 2018 | Application of CP to model the readout of the Genomic Allergen Rapid Detection Assay, used to test skin sensitizers |
| ExCAPE database[8] | Actives *vs* inactives | SVM | QSAR/Classification | Aggregated Mondrian Conformal Prediction | No | Lampa *et al.*[43] | 2018 | Prediction of off-target binding profiles across 31 targets used in early hazard assessment |
| 8 toxicology data sets including off-target functional assays, cytotoxicity tests, mutagenicity tests, CYP450 inhibition assays, acute oral toxicity assays and transporter assays extracted from the literature and public databases | Actives *vs* inactives | RF | QSAR/Classification | Inductive Mondrian Conformal Prediction | No | Ji *et al.*[73] | 2018 | eMolTox: webserver for the prediction of compound toxicity |
| 1,592,127 compounds from ChEMBL version 23 used for model development. The development model was applied to | Compound lipophilicity (logD) | SVM | QSPR/Regression | Cross-Conformal Prediction | No | Lapins *et al.*[74] | 2018 | Large-scale prediction of LogD. The logD prediction model is available as a REST service at https://cplogd.service.pharmb.io/ |

| Data sets (targets; number of compounds; number of datapoints) | Dependent variable or classes | Machine learning technique | Modelling type and task | Type of Conformal Prediction used | Prospective experimental validation? | Reference | Year | Remarks |
|---|---|---|---|---|---|---|---|---|
| 91,498,351 compounds extracted from PubChem database | | | | | | | | |
| 5 data sets from the UCI Machine Learning Repository[75] | Two classes | RF | Binary classification | Aggregated Mondrian Transductive Conformal Prediction | No | Spjuth *et al.*[76] | 2018 | Illustration of a novel framework (Non-Disclosed aggregated Conformal Prediction) to aggregate conformal predictors from multiple sources while keeping the data on which they trained private. High impact for drug discovery research |
| MNIST dataset, and 3 drug discovery data sets measuring the Ames Mutagenicity, mitochondrial function, and interaction with the aryl hydrocarbon receptor of small molecules | Two classes | SVM+ | QSAR/Classification | Inductive Mondrian Conformal Prediction | No | Gauraha et al.[77] | 2018 | Development of inductive conformal predictors using the Learning Under Privileged Information paradigm (LUPI) and SVM+ |
| 550 human protein targets from ChEMBL versions 23 and 24 | pChEMBL | RF | QSAR/Classification | Inductive Mondrian Conformal Prediction | No | Bosc *et al.*[78] | 2019 | Large-scale comparison of QSAR and CP showed that, overall, CP outperforms QSAR, although for some targets the opposite was observed |
| 56,892 compounds from 10 high-throughput | Active/Inactive | Matrix Factorization (Macau) | Multi-target binary classification | Macau + Mondrian Aggregated | No | Norinder *et al.*[80] | 2019 | Multitask learning using Conformal Prediction and matrix factorization |

| Data sets (targets; number of compounds; number of datapoints) | Dependent variable or classes | Machine learning technique | Modelling type and task | Type of Conformal Prediction used | Prospective experimental validation? | Reference | Year | Remarks |
|---|---|---|---|---|---|---|---|---|
| screening data sets originating from PubChem[79] | | | | Conformal Prediction | | | | |
| 24 protein targets from ChEMBL version 23 | $pIC_{50}$ | Fully-connected deep neural networks | Regression | Test-Time Dropout Conformal Prediction | No | Cortés-Ciriano *et al.*[81] | 2019 | First study integrating test-time dropout and Conformal Prediction |

## Conformal Prediction Modalities Commonly Used in Computer-Aided Drug Design

In this section, we will describe the most widely used Conformal Prediction modalities in the drug discovery literature, and the steps required to generate them (see a general pipeline below) using a real-world bioactivity data set. We refer the reader to the work of Vovk *et al*.[32,36,37] for further details about the mathematical foundations underlying the Conformal Prediction framework, and to Norinder *et al.*[32] and Eklund *et al*.[44] for an introduction focused on the application of Conformal Prediction to virtual screening.

The Conformal Prediction implementations we will discuss follow the same core steps, namely:

1. Choosing a non-conformity measure to evaluate the non-conformity between the training and the test instances;
2. Training the machine learning model of choice, and evaluate the non-conformity values for the training examples;
3. Applying the trained model to the test or external set instances;
4. For each test set instance, evaluating its non-conformity with respect to the training data using the same non-conformity measure used in step 1: the higher the conformity of the new instance, the higher the reliability of the prediction;
5. Identifying reliable predictions given the user-defined significance and confidence levels; and
6. Evaluating the validity and efficiency of the generated Conformal Predictor.

### *Inductive Conformal Prediction (ICP)*

### *Inductive Conformal Prediction for Classification*

To generate an ICP model, the data available for training is divided into (see Figure 2) (i) a training set (also termed proper training set[82]), consisting of *e.g.,* 70% of the data, (ii) a calibration set, encompassing *e.g.,* 15% of the data, and (iii) a test set, consisting of the remaining datapoints. Both the proper training and calibration sets are used for training, whereas the test set is used to evaluate the predictive power of the models, as well as the validity and efficiency of the Conformal Predictors generated.

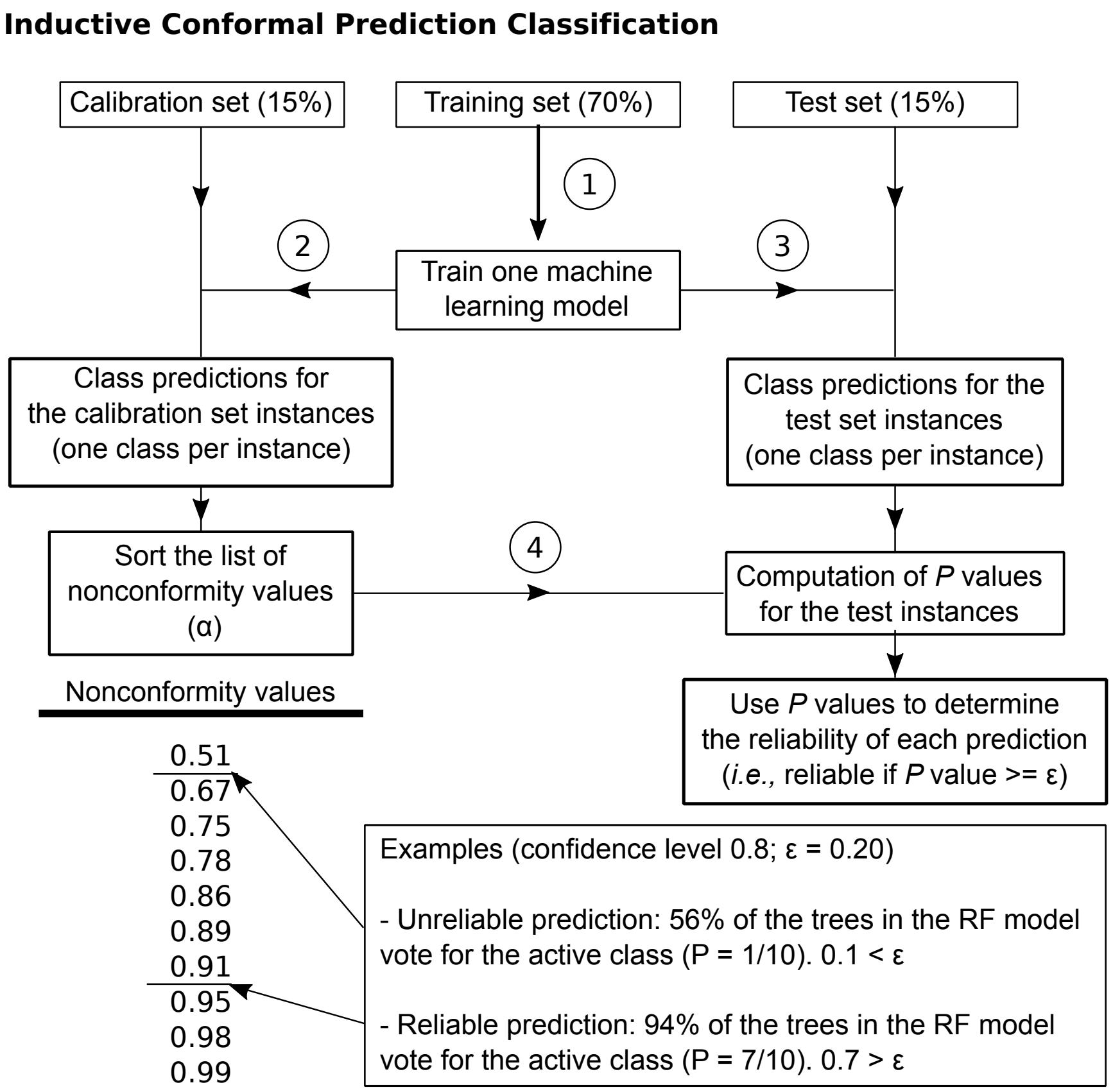

Figure 2. Steps for the generation of an Inductive Conformal Predictor in the context of classification. As in the main text, ε denotes the significance level (Table 1).

A machine learning model is firstly trained on the proper training set (step 1 in Figure 2), and subsequently applied to predict the activity label for the instances in the calibration set (step 2 in Figure 2; for illustration purposes, we here consider that the underlying model is always a RF model unless otherwise stated). The true and predicted class labels for the calibration set instances serve to compute non-conformity scores using the non-conformity measure of choice, for instance, the fraction of Trees from a Random Forest Model voting for the class receiving most votes. The non-conformity score serves to quantify the non-conformity (or ‘strangeness’) of a new instance with respect to those used for training. Predictions for which most of the Trees in the Random Forest predict the same label are more reliable, and hence, have a higher non-conformity score than cases where the algorithm does not delineate both classes correctly; that is, both classes are predicted to be equally likely. The model is then applied to the test set and the fraction of Trees voting for the most voted class is recorded for each instance (step 3 in Figure 2). Finally, the *P* value for each test set instance is calculated as the fraction of instances in the calibration set with a non-conformity score equal or smaller than the nonconformity score for the instance under consideration. Thus, the *P* value represents the ranking of the non-conformity score for a new instance with respect to the non-conformity score list generated for the calibration set. The *P* value is then compared against the significance level, $\varepsilon$, which is defined as 1 – CL. A prediction is considered reliable if the *P* value is higher than $\varepsilon$. Note that the concept of *P* value used here is not equivalent to traditional *P* values used in statistics.

To illustrate this approach using real-world data, we collected $pIC_{50}$ data for 5,207 compounds against the Human ether-a-go-go-related gene (hERG) potassium channel from the ChEMBL database version 23[9]. To generate a RF binary classification model we considered as active those compounds with a $pIC_{50}$ value $\geq$ 7 (n=332), and assigned the remaining compounds to the inactive class (n=4,875). The resulting data set had an imbalance in the ratio of active to inactive compounds of ~1:15. This data set was previously used to benchmark bioactivity modelling pipelines[83], and is publicly available at: https://github.com/isidroc/kekulescope/tree/master/datasets. Next, we calculated circular Morgan fingerprints for all compounds using RDkit (release version 2013.03.02) with a radius of 2 and a fingerprint length of 1,024 bits. To generate an RF-based Inductive Conformal Predictor using a confidence level of 80% we followed the steps described in Figure 2, with the exception that 60% of the data was used as test set in this case for illustration purposes to ensure that we had enough active compounds to compute validity estimates for the resulting Conformal Predictors.

We report in Figure 3 the distribution of non-conformity scores (a) for the calibration and (b) test sets. Unreliable predictions on the test set (shown in red in Figure 3b) are those whose non-conformity value is smaller or equal than the 80$^{th}$ percentile of the list of non-conformity scores for the calibration set, indicated by the cross in Figure 3a, and by the black arrow in Figure 3a. As stated above, the mathematical validity of CP guarantees that at least 80% of the predictions considered reliable will be correct. However, the validity is only guaranteed globally, meaning that the error rate for the reliable predictions (*i.e.,* fraction of predictions that are incorrect) might be different across classes, as some classes are harder to predict than others. This is a major issue when modelling imbalanced data, as the error rate will be higher for, usually, the minority class[39]. In fact, in the hERG data set modeled here we observe that most active compounds are predicted as unreliable (Figure 3c). Therefore, although 80% of the predictions flagged as reliable are correct (global validity), the percentage of active compounds with a reliable prediction is only 24% (28/115; Table 3 and Figure 3), and more importantly, only 14% of the active compounds with a reliable prediction are correctly predicted as active (Table 3). Thus, the local validity for the active class is not guaranteed. This behavior is understandable due to the imbalance of the dataset used, which includes about 15 times more information about inactive compared to active compounds. Being aware of which predictions are more reliable than others now allows the modeler to proceed with those predictions with the desired confidence, *e.g.,* for subsequent experimental compound testing.

| | Reliable predictions | | Unreliable predictions | |
|---|---|---|---|---|
| | Active | Inactive | Active | Inactive |
| Predicted Active | 4 | 0 | 43 | 44 |
| Predicted Inactive | 24 | 1,214 | 44 | 228 |
| Total | 28 | 1,214 | 87 | 272 |

Table 3. Confusion matrix for the classification model trained on the hERG data set, showing that the distribution of reliable and unreliable predictions across classes is not even, and in line with the data distribution in the training data set (which is biased towards the inactive class). The results correspond to the prediction for the test set molecules, and the reliability assignment now allows for the selection of molecules with the desired confidence for subsequent steps of *e.g.*, experimental testing. See also Figure 3.

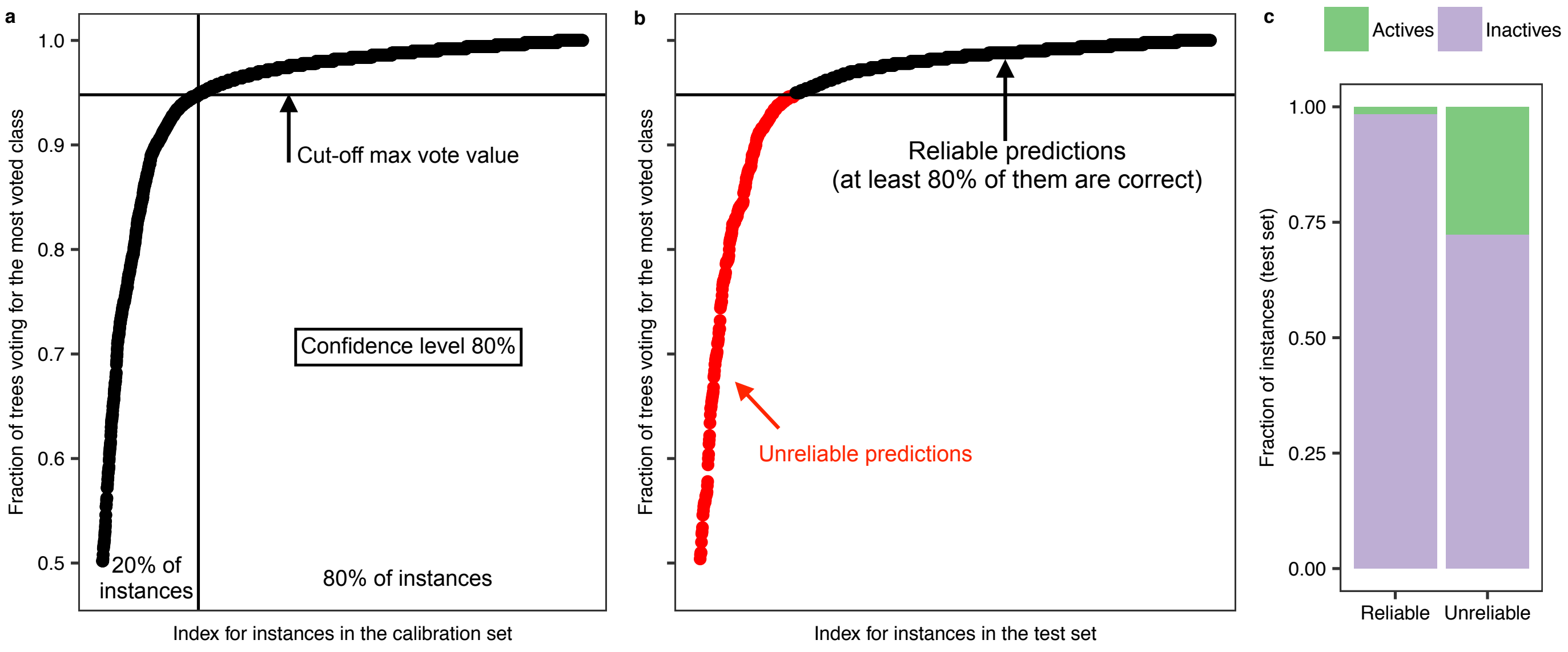

Figure 3 Generation of an Inductive Conformal Predictor using the bioactivity data stored in ChEMBL version 23 for target hERG. The instances in the data set were assigned to the active or inactive class using a cut-off value of 7 $pIC_{50}$ units to generate a highly-imbalanced data set (active to inactive ratio of 1:15). (a) Non-conformity values for the instances in the calibration set. In this example the non-conformity function chosen corresponds to the fraction of Trees voting for the most voted class. For instance, if 85% of the trees in the RF model vote for the active class, the non-conformity score would be 0.85. The sorted list of non-conformity scores serves to calculate $\alpha_{CL.}$ (b) The non-conformity scores are calculated for each instance in the test set. Those predictions with α values equal or greater than $\alpha_{CL}$ are considered reliable. In this case, unreliable predictions are those for which roughly the same number of Trees in the RF model vote for each class. In both (a) and (b) the instances in the *x*-axis have been sorted according to their non-conformity score (*y*-axis). (c) Distribution of inactive and active compounds in the set of reliable (left) and unreliable (right) predictions. The figure shows that most of the reliable predictions correspond to inactive compounds, which are easier to model given the imbalance in the data. Therefore, although at least 80% of the predictions considered reliable are correct, most of the active compounds are assigned an unreliable prediction (Table 3), allowing the modeler to focus on those predictions with the desired confidence in turn.

The lack of validity for each class (or local validity) fostered the development of *Inductive Mondrian Conformal Prediction*[39,84], which permits to calibrate the error rates in a class-specific manner.

### Handling Imbalanced Datasets: Mondrian Conformal Prediction (MCP)

In MCP each class (*e.g.,* active and inactive) is treated separately, and the confidence in the assignment of a given instance to the classes considered is evaluated independently. That is, a list of non-conformity scores is generated for each class using the predictions for the calibration set (Figure 4). Thus, in a binary classification setting a compound might be classified as "active", "inactive", both active and inactive (class "both"), or not assigned to either of them (class "null" or "empty").

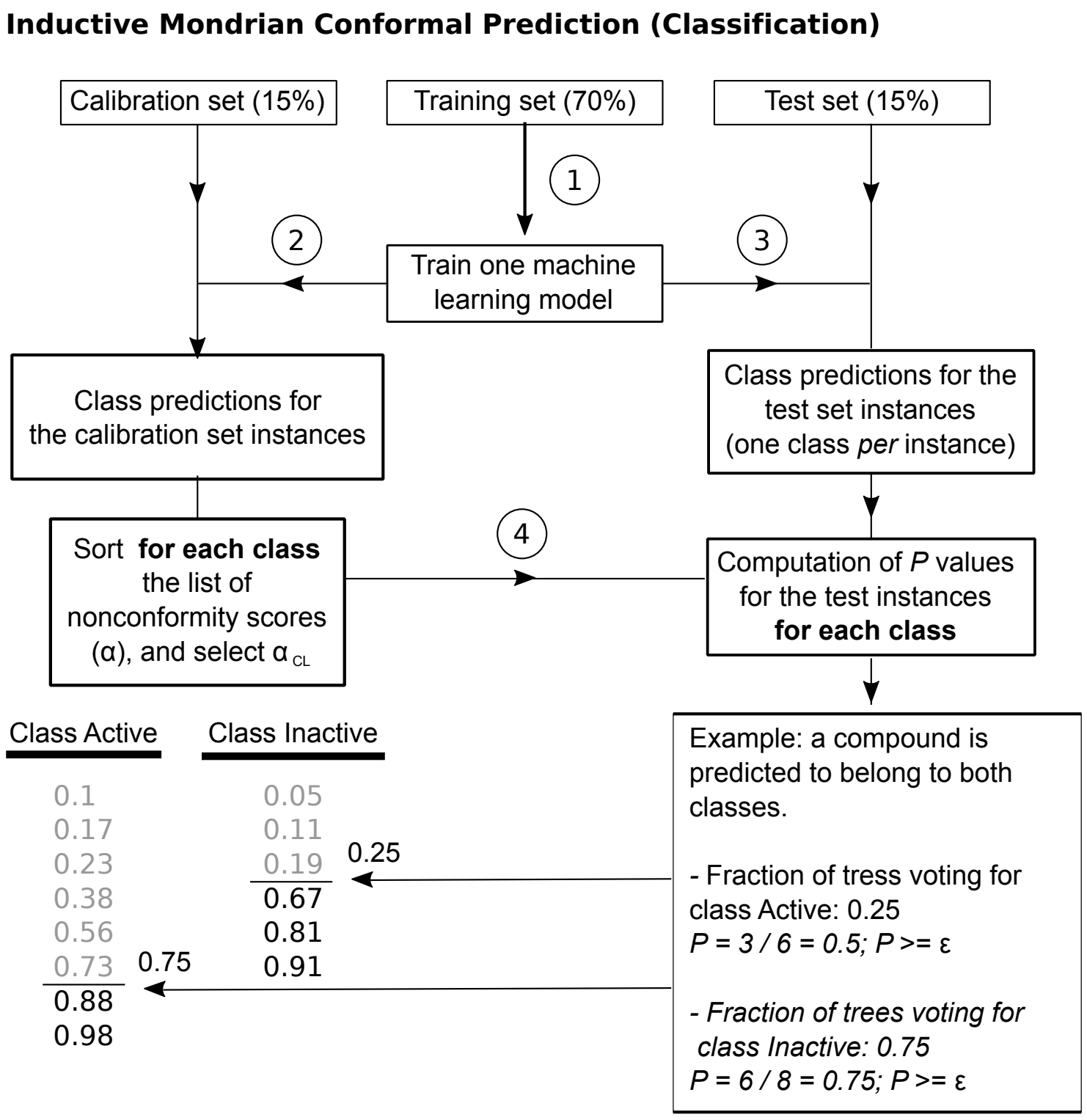

Figure 4. Steps required for the generation of a Mondrian Conformal Predictor.

Specifically, the steps to generate an inductive MCP using a RF classification model are (Figure 4):

- Step 1. Training an RF model on the proper training set;

- Step 2. Applying this model to classify the instances in the calibration set, and creating a list of non-conformity values (*e.g.,* fraction of trees voting for that class in the case of RF) for each category *(i.e.,* active and inactive in a binary classification setting). Next, sorting the lists of non-conformity values in increasing order, which are termed Mondrian class lists;

- Step 3. Applying the model trained in step 1 to the test set, and for each instance compute the fraction of trees voting for each class; and

- Step 4. For each instance in the test set computing a *P* value for each class. The *P* value is the fraction of cases in the corresponding Mondrian class list calculated using the calibration set predictions in Step 3 smaller than the vote fraction for that class. If the *P* value for a given class is above the significance level, ε, the test set instance under consideration is predicted to belong to that category. Compounds are assigned to all categories for which the *P* value is greater or equal than ε.

Hence, a given compound may be predicted as "active", "inactive". However, it can also be predicted as "both" in cases when the model does not have enough predictive power to discriminate between classes, or "null" in cases when the instance is outside the applicability domain of the model. Thus, MCP gives an unbiased estimate of the reliability of the predictions given the training data on a class-specific manner. While this behavior might seem counterintuitive, it is actually a straightforward consequence of the types of data and evidence that might be present in a given training dataset – for example, if there are two closely related molecules to the one a prediction is made for, one of which is active and one of which is inactive, then there is evidence for *both* classes, and (at a given confidence level) none can be chosen over another. Likewise, if no evidence for either class is present in a data set (such as for a chemically very novel molecule), then no prediction either way can be made in practice. This aspect of modelling data is often neglected in other modelling approaches, which forces a decision onto the model, hence resulting in single labels which are often more easy to deal with in practice, but which neither consider the confidence of a prediction properly, nor the underlying evidence of class membership present in the available data.

The significance level, ε, indicates the maximum fraction of predictions that are incorrect. In MCP this fraction is guaranteed for each class, which means that at least 1- ε of the predictions for the minority class will be correct. This is of utmost importance in drug discovery, where for example inactive molecules usually outnumber actives by several orders of magnitude. Likewise, for models in later stages of drug discovery, being able to anticipate for which areas of chemical space no predictions can be made (and to assign a confidence to the remainder) is as important as being able to model the data in the first place.

It is important to note that increasing the confidence level generally reduces efficiency, defined as the single-label prediction rate. In fact, the number of null class predictions is *anticorrelated* with the confidence level, whereas the number of predictions predicted to belong to both categories *positively correlates* with the significance level[32]. This relationship between confidence level and one-class assignments is illustrated in Figure 5, where we show the results generated using a MCP trained on the hERG data set described above (see also Table 4 and Figure 3). Hence in practice the modeler needs to make choices as to which confidence level and efficiency is desired in a particular case, which involves a certain amount of subjective choice as well.

Since the introduction of MCP in the chemical-structure activity modelling community[32], the advantages of MCP have been showcased in a number of applications, mostly to perform classification tasks using imbalanced data sets, which is the situation MCP sets out to address. Norinder *et al*.[32] applied MCP to binary classification of compounds and showed that the number of null predictions and the confidence level are inversely correlated, whereas the opposite is observed for the *both* category (Figure 6). Therefore, increasing the confidence might lead to assign molecules to the both category that are already assigned only to the correct category at a lower confidence level, underlining that this parameter represents a trade-off between multiple factors, and that a higher confidence level is not always the better choice (Figure 6). By modelling Ames mutagenicity data using Mondrian ICP Norinder *et al.*[6] showed that the inconsistency across data sets from different sources seems to prevent the generation of valid conformal predictors, likely in this case due to inconsistencies in categorizing moderately mutagenic compounds as mutagenic or non-mutagenic, which highlights the importance of evaluating data consistency prior to modelling[85–87]. This can also be seen as a useful feature of CP algorithms, where the model itself is able to detect how consistent training data is which it is trained with, in the same way it is able to assign a confidence level to the output values. Norinder *et al.*[4] also implemented Inductive MCP using Random Forest as the underlying algorithm to model the mutagenicity of 936 primary aromatic amines (630 mutagenic and 306 nonmutagenic) using Leadscope fingerprints[88] where it was found that models were valid for both classes. A recent large-scale comparison of QSAR and MCP models for ligand-target prediction using *in vitro* activity data for 550 human protein targets extracted from ChEMBL found that the predictive power of both approaches (in terms of correct classification rate) is overall similar for 92% of the targets considered at a confidence level of 80%[78]. The usefulness of MCP as a robust method to determine the applicability domain of predictive models for regulatory purposes has been shown by Norinder *et al.*[39] using carcinogenicity and mutagenicity data. Overall, these studies showcase the versatility of MCP to handle imbalanced data sets. However, holding out the calibration set to generate the list of non-conformity scores hampers the use of all available labelled data for model training. Therefore, several flavours of CP designed to use all available data have been developed, which we will revisit later in this review (see section 'Conformal Prediction Using All Labelled Data for Learning').

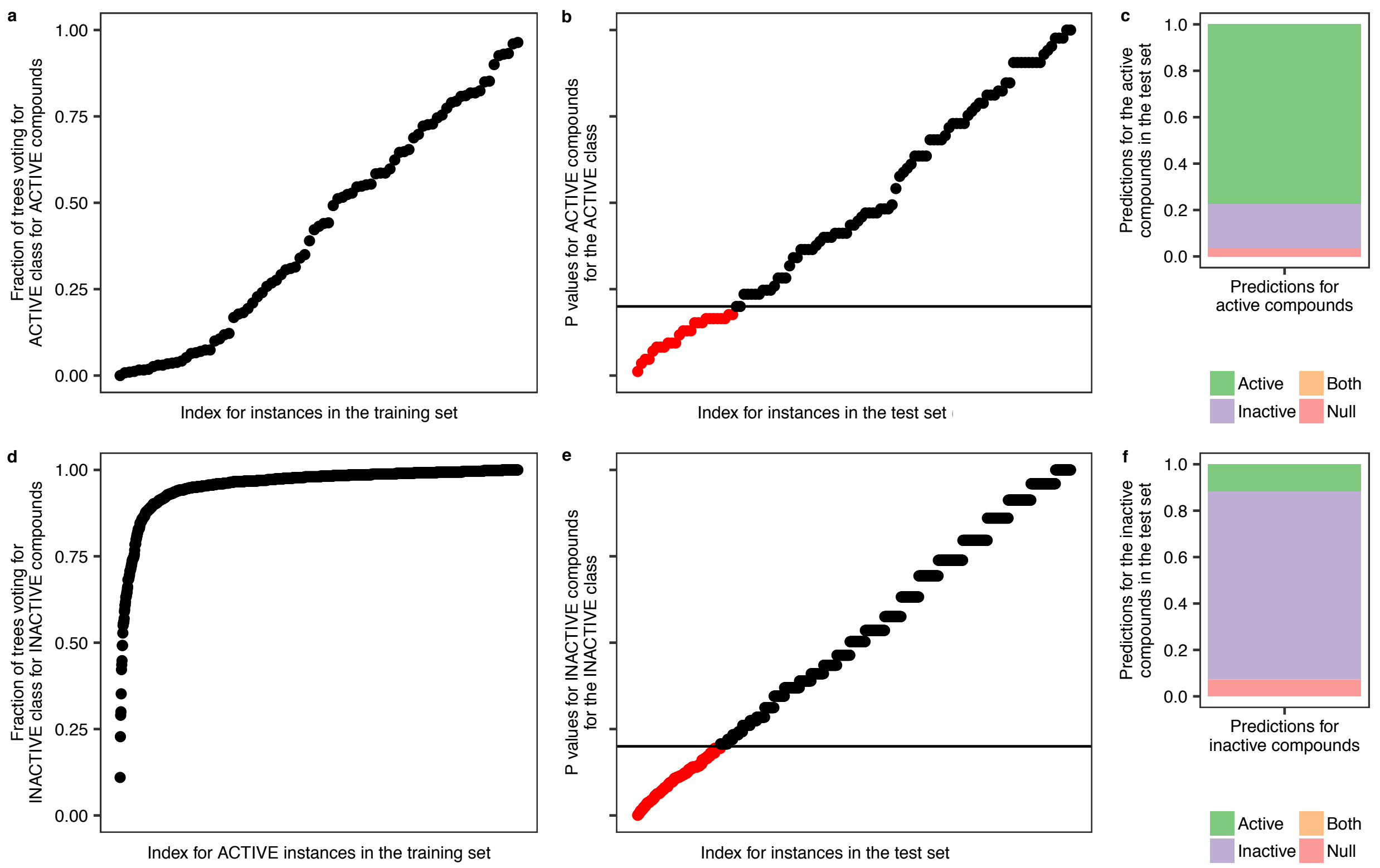

Figure 5. Modelling the hERG data set using Mondrian Conformal Prediction. (a) Fraction of trees voting for the active class for the active compounds in the calibration set. (b) Fraction of trees voting for the active class for the active compounds in the test set. Unreliable predictions (*i.e.,* those with *P* values below the confidence level selected, 80%) are shown in red. (c) Distribution of class predictions for the active compounds in the test set. (d) Fraction of trees voting for the inactive class for the inactive compounds in the calibration set. (b) Fraction of trees voting for the inactive class for the inactive compounds in the test set. Unreliable predictions are shown in red. (c) Distribution of class predictions for the active compounds in the test set. It can be seen that the fraction of instances whose try value is among the set predicted labels is ~0.8 for both categories, which corrresponds to the selected confidence level in this case. The instances in a, b, d, and e have been sorted in increasing order according to the value of their associated non-conformity score (*y*-axis).

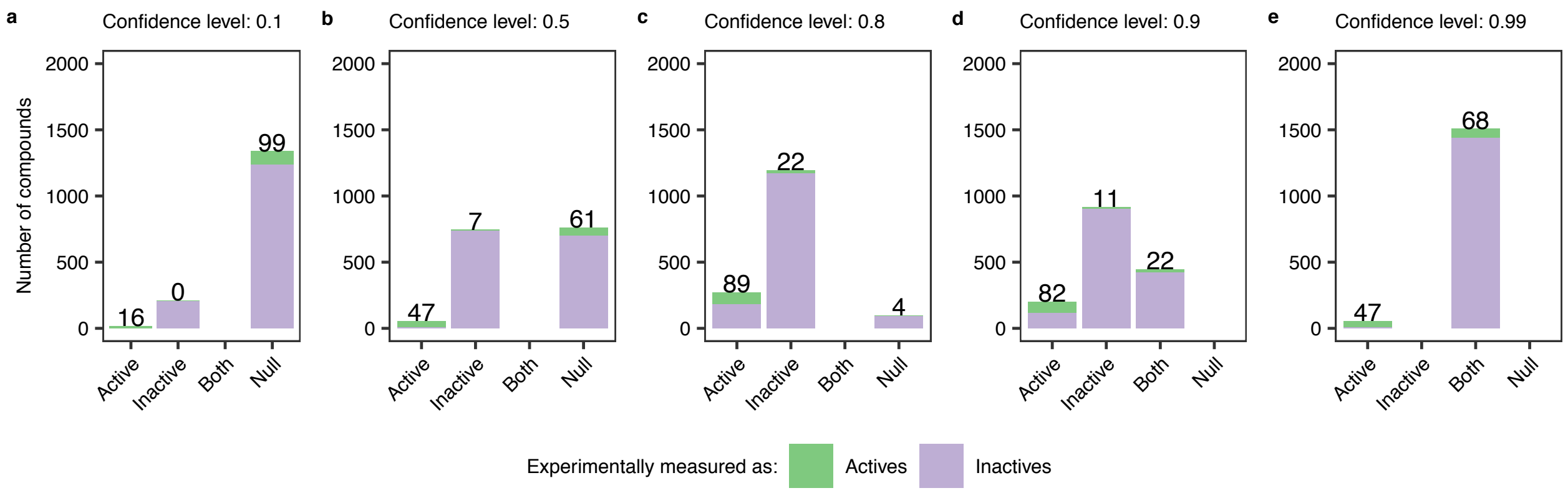

Figure 6. Influence of the confidence level on the efficiency of MCP classification models trained on the hERG data set and using increasingly larger confidence levels: 0.1 (a), 0.5 (b), 0.8 (c), 0.9 (d), and 0.99 (e). The numbers on top of the bars indicate the number of active compounds in each set. The *x*-axis indicates the predicted categories, whereas the colours indicate the true category for the compounds (*i.e,* active or inactive). It can be seen that the number of instances predicted to belong to both classes (category 'both') increases with the confidence level.

Table 4. Performance of the MCP model trained for the hERG data set. It can be seen that the predictions are globally *and* locally valid*, i.e.* across the model and also for individual classes.

| Confidence level | Global validity | Validity for actives | Validity for inactives |
|---|---|---|---|
| 0.1 | 0.14 | 0.14 | 0.14 |
| 0.5 | 0.5 | 0.41 | 0.51 |
| 0.8 | 0.81 | 0.77 | 0.81 |
| 0.9 | 0.92 | 0.9 | 0.92 |
| 0.99 | 0.99 | 1 | 0.99 |

### *ICP for Regression*

The underlying principles to generate Inductive Conformal Predictors for regression tasks are similar to those described above for classification models. Firstly, a model is trained on the proper training set (step 1 in Figure 7). Subsequently, the model is applied to the calibration set (step 2 in Figure 7). In the case RF-based ICP models the predicted value for each instance in the validation set, $\hat{y}$, is then calculated as the average across the Trees in the Random Forest, and the standard deviation of these, $\sigma$, is used as a measurement of the prediction's uncertainty[89]. Scaling the absolute error in prediction using a measure of confidence about each prediction (*e.g.,* the bagged variance[29]) serves to generate tighter predictions for those prediction that are deemed more reliable[90]. Note that without this scaling all predictions for the new molecules would be of the same size (Equation 1).

The residuals and the standard deviation across the forest are used to generate a list of non-conformity scores for the calibration set as follows:

$$\text{Equation 1:} \quad \propto_i = \frac{|y_i - \hat{y}_i|}{e^{\sigma_i}}$$

where $y_i$ is the $i^{th}$ instance in the validation set, and $\hat{y}_i$ and $\sigma_i$ are the average and the standard deviation of the predicted activities for the $i^{th}$ instance across the forest, respectively. The resulting list of non-conformity scores, $\alpha$, is sorted in increasing order, and the percentile corresponding to the confidence level considered is selected, *e.g.,* the 80$^{th}$ percentile for a confidence level of 0.80 ($\alpha_{80}$). Next (step 4 in Figure 7), the standard deviation across the forest is used to calculate confidence regions for the data points in the test set as follows (step 4 in Figure 1):

$$\text{Equation 2:} \quad Confidence\ region = \hat{y}_j \pm |y_j - \hat{y}_j| = \hat{y}_j \pm (e^{\sigma_j} * \alpha_{CL})$$

Where $y_j$ is the $j^{th}$ instance in the test set, $\hat{y}_j$ and $\sigma_j$ are the average and the standard deviation of the predicted activities for the $j^{th}$ instance across the forest, respectively, and $\alpha_{CL}$ is the non-conformity score for the selected confidence level.

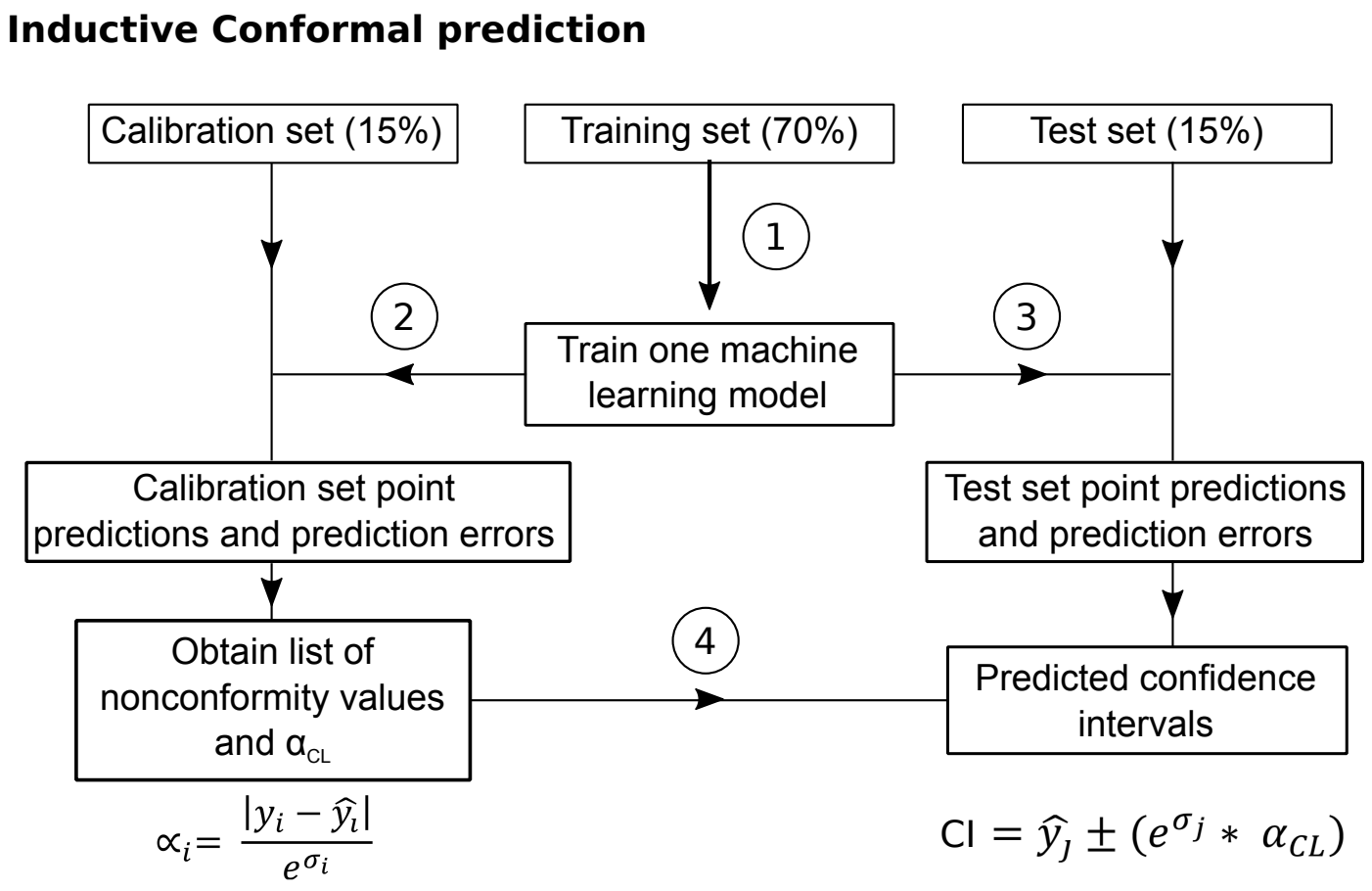

Figure 7. Steps for the generation of Inductive Conformal Predictors in the context of regression.

Choosing an appropriate non-conformity measure is essential in regression to maximize efficiency, defined as the average size of the confidence regions. Svensson *et al.*[40] benchmarked the efficiency of the following 6 approaches to scale the errors in prediction in the non-conformity function (Equation 1) using RF models and 29 public bioactivity data sets: (i) standard deviation across the trees in the RF model, (ii) the interpercentile range ($10^{th}$ to $90^{th}$) of the predictions across the trees in the RF model, (iii) the predicted error generated by a separate model trained on compound descriptors and using the cross-validation residuals as the dependent variable, (iv) the distance in two-component Principal Component Analysis (PCA) space computed using compound descriptors to the center of all training data, (v) the average distance to the five nearest neighbours in two-dimensional incremental PCA, and (vi) t-SNE space. Although the 6 non-conformity functions generated well-calibrated confidence intervals, their efficiency varied considerably, showing that the natural exponential of the RF ensemble standard deviation led to the tightest intervals, with an average prediction range of 1.65 $pIC_{50}$ units; using the ensemble interpercentile range and a separate error model led to slightly worse performance. The exponential scaling sets the upper value for the list of non-conformity scores to be equal to the largest residual in the calibration set, as the exponential converts low $\sigma$ to values close to unity[40,63,81]. Lapins *et al*[74]. included a smoothing factor to (i) make large intervals slightly smaller, thus mitigating the effect of potentially inaccurate, large $\sigma$ values, and (ii) increase the size of very small intervals not reflecting the inherent uncertainty of bioactivity measurements[85–87,91].

A number of the Conformal Predictors reported in the literature applied to regression tasks use RF models as the underlying algorithm (Table 2), and employ the standard deviation across base learners to scale the residuals and compute non-conformity scores[40,92] (Equation 1), a choice supported by comparative analysis of non-conformity measures[40]. RF models are widely used in Conformal Prediction because the generation of an ensemble comes at no extra computational cost[15,32], the training can be parallelized, and the performance is stable across parameter values, thus requiring little parameter tuning. The variance across base learners (*i.e.,* the bagged variance across the trees in RF models) conveys a predictive signal to quantify the uncertainty of individual predictions, as the average RMSE on the test set increases with the variance among predictions[22,29]. However, one should note that this numerical Pearson correlation between variance and prediction error is much weaker than the inflated correlation obtained by binning the test compounds on the basis of the predicted variance[22,29,93], which in practice means that the size of the confidence regions calculated using non-conformity measures based on the bagged variance will not be strongly correlated with the unsigned error in prediction. In other words, the average RMSE on the test set is correlated with the size of the confidence interval computed using Conformal Predictors because, *on average* (but not in every individual case) the larger the variance across the ensemble, the larger the predicted confidence region will be (Equation 2). Other studies have used a second machine learning model, termed error model, to predict errors in prediction to identify which predictions are less reliable in a similar manner as the standard deviation across the forest is used in RF-based conformal predictors (Equation 1). This is achieved by training a model on the same chemical descriptors used to train a point prediction model to predict the error in prediction during cross-validation[92] or for the calibration set instances[32]. In other cases other metrics are used as covariates[31,94], or linear methods[95]. Overall, this approach has been shown to deliver less efficient Conformal Predictors as compared to those generated using the bagged variance when modelling QSAR data sets[40].

Inductive CP has been applied to diverse regression tasks (Table 2), including proteochemometric modelling of PARP inhibitors[46] and the prediction of patterns of growth inhibition across cancer cell line panels[92]. However, most studies using CP for regression tasks have implemented CP modalities designed to use all available training data for learning, which we discuss in the next section.

### *Conformal Prediction Using All Labelled Data for Learning*

A common disadvantage to all the CP modalities revisited so far is that not all data available for training are used for model fitting, as the calibration set needs to be kept aside to generate the list of non-conformity scores (Figures 2 and 4). However, it would be desirable to use all available data during training to increase the predictive power of the resulting models, in particular in cases where few data points are available for a given class or a particular region of the bioactivity range considered. Several CP modalities have been developed to date to solve this issue[96,97]. We here revisit the two most widely used in drug discovery in both regression and classification tasks (Table 2), namely Aggregated Conformal Prediction (ACP)[98,99] and Cross Conformal Prediction (CCP)[96].

- **Aggregated Conformal Prediction (ACP)**

The ACP approach consists of generating a collection of usually 10-100[43,56,80] Inductive Conformal Predictors (*e.g.,* Inductive MCP), each of them trained on the same training data but with random assignments of training set instances to the proper training and calibration sets. Thus, the instances assigned to the calibration and proper training sets are different for each model, resulting in reduced variance for the predicted confidence regions, and increased efficiency[98]. In the case of classification, each model is applied to the test (or external) set instances, and the *P* values computed with each of these models are recorded for each instance (and for each class in the case of MCP). The final *P* value is calculated as *e.g.,* the median *P* value and each instance is assigned to those classes for which the median *P* value is higher than the significance level. In the case of regression, the final confidence interval for each test set instance is also a function of the set of confidence intervals predicted by each of the set of models trained, such as half the difference between the median of the maximum and minimum predicted values[61].

ACP has been applied in diverse drug discovery tasks (Table 2), including Ames mutagenicity prediction[67], multi-task learning using matrix factorization[80], modelling of compound binding to the estrogen and androgen receptors[54], and virtual screening of TRPV1 agonists[100]. Lindh *et al.*[61] generated Aggregated Conformal Predictors to model the permeation rate through the human skin of 211 chemical compounds[60]. Whereas the predictive power did not increase with respect to previous models reported for the same data set, CP added the advantage of providing predictions as confidence regions, and using all available data for training, which was crucial in this particular case due to the limited size of the training data. ACP has also proved versatile to model highly imbalanced data sets. For instance, Svensson et al.[70] modelled 12 highly-imbalanced bioactivity data sets from PubChem using ACP and a cost-gain function taking into account screening costs and the gain of finding active hits. ACP was also shown to accurately model highly-imbalanced cytotoxicity data sets from PubChem[62], with only 0.8% of cytotoxic compounds, showing an external validation set a sensitivity of 74% and a specificity of 65% for the single-label predictions at a confidence level of 80%. Lastly, ACP has also been applied in two iterative screening studies, where Conformal Prediction followed molecular docking in order to prioritize compounds for experimental testing[56,101]. Overall ACP has been found to reduce the variance and increase the efficiency of the predicted confidence intervals, while also permitting the use of all available data for training.

- **Cross-Conformal Prediction (CCP)**

In CCP the training data set is divided into *k* non-overlapping sets, in a similar manner as performed in *k*-fold cross validation. Next, *k* ICP models are trained, each time using a different set as calibration set, thus permitting the use of all labelled data for training. For each instance in the test set *k P* values are generated, whose *e.g.,* mean value can be used as the output *P* value. The output *P* values are compared against the significance level to assign a class to the test instances (or classes in the case of Mondrian CCP). To date, Mondrian CCP has been mostly applied to model imbalanced bioactivity data sets (Table 2). Sun *et al.*[57] reported SVM-based classification Mondrian CCP models showing comparable performance to SVM models on external sets for 18 data sets from the ExCAPE database[8] with active to inactive ratios in the 1:10 to 1:1000 range. The Mondrian CCP models were well-calibrated for both the majority and the minority class. Notably, the authors reported higher efficiency for Mondrian CCP models using a non-conformity measure based on the distance to the SVM decision boundary as compared to a non-conformity measure based on the Tanimoto distance to the 5 nearest neighbours in the training data. Giblin *et al.*[69] conducted one of the few studies where CP guided prospective validation experiments. Specifically, the authors built Mondrian CCP proteochemometric[102] models using bioactivity data for 6,352 compounds across 31 bromodomains (15,350 data points), which showed a maximum Matthew's Correlation Coefficient of 0.83 on an external test set. 319 compound-target pairs were confirmed out of 1,139 experimentally validated, where the selection included compounds with high and low confidence values and different bioactivity profiles against sets of bromodomains. In the context of regression, CCP has been applied to model compound lipophilicity on a large scale. Lapins *et al.*[74] harnessed calculated logD values for 1,592,127 compounds extracted from ChEMBL version 23 to generate an SVM-based CCP using 10-fold cross validation. The errors in prediction were calculated as the median prediction midpoint +/- the median predicted interval size across the 10 models. The final model, which achieved median confidence intervals of ± 0.39 and 0.60 log units at 80% and 90% confidence, respectively, was further applied to 91,498,351 compounds extracted from the PubChem database. These predictions were made publicly available at https://cplogd.service.pharmb.io. Overall, these studies show the versatility of CCP prediction, which permits to obtain valid and efficient conformal predictors while using all available labelled data for training.

**Conformal Prediction Methods for Deep Learning**

Deep learning is currently applied in many tasks of the drug discovery process, a trend that is only expected to rise in the next years[103,104]. The predictions output by deep learning classification models are overall well calibrated, or can be easily calibrated using *e.g.,* Platt Scaling or Isotonic Regression[105]. Therefore, there is little benefit in using CP on top of deep learning models generated for classification tasks. However, in the case of regression deep learning models simply output point predictions, thus not providing information about the reliability of individual predictions directly. In addition to variational Bayesian inference methods[20,21], CP has been applied to deep learning models to estimate reliable confidence intervals for individual predictions. The application of CP to regression networks was initially proposed by Papadopoulos *et al.*[95]. In drug discovery, the authors recently proposed two methods to obtain reliable confidence intervals for regression networks at minimal computational cost. The first approach, Deep Confidence[63], harnesses the predictions generated by intermediate network states corresponding to the local minima visited during the training of a single network to build an ensemble of predictions. The variability across the ensemble can be used to scale the absolute errors in prediction in a similar way as performed using the bagged variance in the case of RF models. More recently, we proposed a second method, Test-Time Dropout Conformal Prediction[81], that consists of training a network using dropout[106]. Next, the network is applied to compute *N* forward passes using dropout as well. As in the case of Deep Confidence, the variability across these forward passes is used to scale the absolute errors in prediction. Overall, it could be shown that both Deep Confidence and Test-Time Dropout Conformal Prediction deliver well-calibrated predicted confidence intervals that, in addition, span a narrower range of values than those computed using RF-based models.

**Open-Source Implementations of Conformal Prediction**

The increased adoption of CP in early-stage drug discovery settings has fostered the implementation of several CP modalities in open-source software libraries in the R programming language and Python, which are widely used programming languages in medicinal chemistry applications[107,108]. The availability of predictive modelling packages in both R (*e.g.*, *caret*[109] or *camb*[110]) and Python (*e.g., scikit-learn*[111] or *PyTorch*[112]) has facilitated the integration of existing computational drug discovery pipelines and CP[56,110].

The Python *nonconformist* package (http://donlnz.github.io/nonconformist/index.html) provides functionalities to generate Inductive and Aggregated Conformal Predictors for both regression and classification tasks. CPSign[113] (http://cpsign-docs.genettasoft.com/) is another Python implementation of CP for chemoinformatic tasks that uses SVM and Signature molecular descriptors. The R package *conformal*[46] (https://github.com/isidroc/conformal) is an object-oriented programming implementation of CP that provides *S4* classes to construct classification and regression ICP models using Random Forest as the underlying algorithm. Similarly, the R package *conformalClassification* (https://cran.r-project.org/web/packages/conformalClassification/index.html) permits to generate Transductive and Inductive Conformal Predictors based on RF models for classification tasks. The availability of well-documented and structured software will increase the reproducibility of published results, and allow for robust benchmarking studies of novel methods. Hence, we advocate for the publication of source code in future CP studies, in line with current recommendations in the modelling community[114,115].

**Current Limitations of Conformal Prediction and Future Perspectives**

While Conformal Prediction, as described above, is able to assign confidence to predictions within a computationally efficient framework some areas of ongoing methodological research certainly remain. A major issue in CP applied to regression is the low efficiency of most CP models, which leads to the predicted confidence regions to span multiple *e.g.*, $pIC_{50}$ units. Such large intervals are not informative and thus hamper the practical usefulness of CP. Substantial efforts in the community have been invested in investigating and developing non-conformity functions to reduce the size of the predicted confidence regions[40,63,90,116]. However, future research will be needed to improve current non-conformity measures in order to, ultimately, generate errors in prediction comparable to the uncertainty of the data[85–87]. In classification settings, a common problem faced is the substantial increase of instances predicted to belong to multiple categories as the confidence level is increased[32,78]. This problem is analogous to the lack of efficiency of regression models. As in the regression case, the development of improved non-conformity measures[57,80] will be needed to improve the efficiency of classification CP models. Another major shortcoming of the predicted confidence regions in the case of regression is the poor correlation between the absolute error in prediction (*i.e.,* unsigned error) and the size of the confidence interval. This is due to the fact that error models used in non-conformity measures *e.g.,* the bagged variance, do not predict accurately the error in prediction[89,94] . Thus, large confidence intervals are obtained for accurate predictions and *vice versa.* Using alternative methods to the bagged variance to compute nonconformity scores, such as quantitative metrics developed in the QSAR arena to estimate the applicability domain of the models[26], might alleviate this issue. Similarly, algorithms other than the most widely used to date (RF, SVM and neural networks; Table 2) might also be considered in drug discovery applications as alternative methods to generate more efficient Conformal Predictors[80,90].

Today, Mondrian CP modalities, including Mondrian ACP and CCP, have become the standard approach to model imbalanced data sets when using Conformal Prediction. However, also the data sets themselves which are used to model compound activity on a continuous scale are generally biased, which leads to an uneven coverage of the chemical space across the bioactivity range considered, and hence variable errors rates across it[94]. Therefore, the development of methods to handle the uneven distribution of datapoints across the bioactivity range considered in regression models would be useful to remove biases from models, and hence guarantee that the validity and efficiency of the predicted confidence intervals are even across the bioactivity range modelled, and not only across the entire bioactivity range.

The integration of predictions generated by independent conformal predictors is a current area of intense research, similar to ensemble approaches in other domains. Toccaceli et al.[117] recently introduced a method to integrate ICP models trained on different underlying algorithms. Notably, the combined models outperformed base learners (linear SVM, Gradient Boosted Trees, and *k*-Nearest Neighbours) on an IDH1 bioactivity data set extracted from the ExCAPE database[8]. An alternative approach to integrate CP models, applicable to both classification and regression models, is Synergy CP[97], which permits the aggregation of CP models trained in parallel on subsets of the training data into valid and efficient Conformal Predictors. Overall, these studies represent innovative solutions that will permit not only performance improvements, but also the exploitation for drug discovery of (proprietary) data dispersed across companies and institutions in distributed environments.

As stated above, the validity of CP is only guaranteed if the randomness or exchangeability assumptions hold. This assumption is however not usually verified in practice, and it is only assumed that the training data and the molecules to which CP models are applied are drawn from the same distribution. It is of course unlikely that the chemical space covered in the training data of virtual screening models, even if these encompass thousands of molecules, is entirely representative of the entire chemical space already comprised in academic and commercial chemical libraries, or amenable to chemical synthesis. In fact, the authors recently showed using iterative virtual screening experiments that breaching the randomness assumption leads to useless conformal predictors[41]. This issue is of particular relevance given that many of the CP reported to date are based on few hundred datapoints (Table 2). Therefore, further development of methods to determine to what extent the randomness or exchangeability assumptions hold would be useful in practice to make informed decisions on the applicability of the developed CP models on the basis of the difference between the training data and those molecules to which the models are applied.

**Acknowledgements**

This project has received funding from the European Union’s Framework Programme For Research and Innovation Horizon 2020 (2014-2020) under the Marie Curie Sklodowska-Curie Grant Agreement No. 703543 (I.C.C.).

**Conflicts of Interest**

The authors declare no conflict of interests.

## References

1. Vayena, E., Blasimme, A. & Cohen, I. G. Machine learning in medicine: Addressing ethical challenges. *PLOS Med.* **15**, e1002689 (2018).
2. Segall, M. D. & Champness, E. J. The challenges of making decisions using uncertain data. *J. Comput. Aided. Mol. Des.* **29**, 809–16 (2015).
3. Hanser, T., Barber, C., Marchaland, J. F. & Werner, S. Applicability domain: towards a more formal definition. *SAR QSAR Environ. Res.* **27**, 865–881 (2016).
4. Tetko, I. V, Bruneau, P., Mewes, H.-W., Rohrer, D. C. & Poda, G. I. Can we estimate the accuracy of ADME-Tox predictions? *Drug Discov. Today* **11**, 700–707 (2006).
5. Weaver, S. & Gleeson, M. P. The importance of the domain of applicability in QSAR modeling. *J. Mol. Graph. Model.* **26**, 1315–1326 (2008).
6. Toplak, M. *et al.* Assessment of Machine Learning Reliability Methods for Quantifying the Applicability Domain of QSAR Regression Models. *J. Chem. Inf. Model.* **54**, 431–441 (2014).
7. Netzeva, T. I. *et al. Current Status of Methods for Defining the Applicability Domain of ( Quantitative ) Structure – Activity Relationships*. **2**, (2005).
8. Sun, J. *et al.* ExCAPE-DB: an integrated large scale dataset facilitating Big Data analysis in chemogenomics. *J. Cheminform.* **9**, 17 (2017).
9. Gaulton, A. *et al.* The ChEMBL database in 2017. *Nucleic Acids Res.* **45**, D945–D954 (2017).
10. Cherkasov, A. *et al.* QSAR modeling: where have you been? Where are you going to? *J. Med. Chem.* **57**, 4977–5010 (2014).
11. Lecun, Y., Bengio, Y. & Hinton, G. Deep learning. *Nature* **521**, 436–444 (2015).
12. Todeschini, R. & Consonni, V. *Handbook of Molecular Descriptors*. (2008). doi:10.1002/9783527613106
13. Rivas-Perea, P. *et al.* Support Vector Machines for Regression: A Succinct Review of Large-Scale and Linear Programming Formulations. *Int. J.* **3**, (2013).
14. Cortes, C. & Vapnik, V. Support-Vector Networks. *Mach. Learn.* **20**, 273–297 (1995).
15. Breiman, L. Random Forests. *Mach. Learn.* **45**, 5–32 (2001).
16. Obrezanova, O., Csányi, G., Gola, J. M. R. & Segall, M. D. Gaussian Processes: A Method for Automatic QSAR Modeling of ADME Properties. *J. Chem. Inf. Model.* **47**, 1847–1857 (2007).
17. Cortes-Ciriano, I. *et al.* Proteochemometric modeling in a Bayesian framework. *J. Cheminf.* **6**, 35 (2014).
18. Obrezanova, O. & Segall, M. D. Gaussian processes for classification: QSAR modeling of ADMET and target activity. *J. Chem. Inf. Model.* **50**, 1053–1061 (2010).
19. Rasmussen, C. E. & Williams, C. K. I. *Gaussian Processes for Machine Learning*. (Mit Press, 2006).
20. Zhang, Y. & Lee, A. A. *Bayesian semi-supervised learning for uncertainty-calibrated prediction of molecular properties and active learning*. (2019).
21. Gal, Y., Ghahramani, Z., Uk, Z. A. & Ghahramani, Z. Dropout as a Bayesian Approximation: Representing Model Uncertainty in Deep Learning. *2015, arXiv1506.02142 arXiv.org ePrint Arch. https//arxiv.org/abs/1506.02142*

*(accessed Jul 10, 2018).* (2015).
22. Sheridan, R. P. Using Random Forest To Model the Domain Applicability of Another Random Forest Model. *J. Chem. Inf. Model.* **53**, 2837–2850 (2013).
23. Svetnik, V. *et al.* Random forest: a classification and regression tool for compound classification and QSAR modeling. *J. Chem. Inf. Comp. Sci.* **43**, 1947–1958 (2003).
24. Berenger, F. & Yamanishi, Y. A Distance-Based Boolean Applicability Domain for Classification of High Throughput Screening Data. *J. Chem. Inf. Model.* **59**, 463–476 (2019).
25. Liu, R. & Wallqvist, A. Molecular Similarity-Based Domain Applicability Metric Efficiently Identifies Out-of-Domain Compounds. **59**, 181–189 (2019).
26. Netzeva, T. I. *et al.* Current status of methods for defining the applicability domain of (quantitative) structure-activity relationships. The report and recommendations of ECVAM Workshop 52. *Altern. Lab. Anim.* **33**, 155–173 (2005).
27. Sushko, I. *et al.* Applicability domain for in silico models to achieve accuracy of experimental measurements. *J. Chemom.* **24**, 202–208 (2010).
28. Sahigara, F. Defining the Applicability Domain of QSAR models : An overview. *Mol. Descriptors. Free online Resour.* 1–6 (2007).
29. Wood, D. J., Carlsson, L., Eklund, M., Norinder, U. & Stålring, J. QSAR with experimental and predictive distributions: an information theoretic approach for assessing model quality. *J. Comput. Aided Mol. Des.* **27**, 203–219 (2013).
30. Schroeter, T. S. *et al.* Estimating the domain of applicability for machine learning QSAR models: a study on aqueous solubility of drug discovery molecules. *J. Comput. Mol. Des.* **21**, 485–498 (2007).
31. Sheridan, R. P. The Relative Importance of Domain Applicability Metrics for Estimating Prediction Errors in QSAR Varies with Training Set Diversity. *J. Chem. Inf. Model.* **55**, 1098–1107 (2015).
32. Norinder, U. *et al.* Introducing Conformal Prediction in Predictive Modeling. A Transparent and Flexible Alternative To Applicability Domain Determination. *J. Chem. Inf. Model.* **54**, 1596–1603 (2014).
33. Schwaighofer, A. *et al.* Accurate solubility prediction with error bars for electrolytes: a machine learning approach. *J. Chem. Inf. Model.* **47**, 407–424 (2007).
34. Liu, R., Glover, K. P., Feasel, M. G. & Wallqvist, A. General Approach to Estimate Error Bars for Quantitative Structure–Activity Relationship Predictions of Molecular Activity. *J. Chem. Inf. Model.* **58**, 1561–1575 (2018).
35. Cortes-Ciriano, I., Murrell, D. S., van Westen, G. J. P., Bender, A. & Malliavin, T. Prediction of the Potency of Mammalian Cyclooxygenase Inhibitors with Ensemble Proteochemometric Modeling. *J. Cheminf.* **7**, 1 (2014).
36. Vovk, V., Gammerman, A. & Shafer, G. *Algorithmic learning in a random world*. (Springer, 2005).
37. Shafer, G. & Vovk, V. A Tutorial on Conformal Prediction. *J. Mach. Learn. Res.* **9**, 371–421 (2008).
38. Vovk, V., Fedorova, V., Nouretdinov, I. & Gammerman, A. Criteria of efficiency for conformal prediction. in *Lecture Notes in Computer Science (including subseries Lecture Notes in Artificial Intelligence and Lecture Notes in Bioinformatics)* **9653**,

23–39 (2016).
39. Norinder, U., Carlsson, L., Boyer, S. & Eklund, M. Introducing conformal prediction in predictive modeling for regulatory purposes. A transparent and flexible alternative to applicability domain determination. *Regul. Toxicol. Pharmacol.* **71**, 279–284 (2015).
40. Svensson, F. *et al.* Conformal Regression for Quantitative Structure–Activity Relationship Modeling—Quantifying Prediction Uncertainty. *J. Chem. Inf. Model.* **58**, 1132–1140 (2018).
41. Cortes-Ciriano, I., Firth, N. C., Bender, A. & Watson, O. Discovering highly potent molecules from an initial set of inactives using iterative screening. *J. Chem. Inf. Model.* **58**, 2000–2014 (2018).
42. Johansson, U., Linusson, H., Löfström, T. & Boström, H. Interpretable regression trees using conformal prediction. *Expert Syst. Appl.* **97**, 394–404 (2018).
43. Lampa, S. *et al.* Predicting Off-Target Binding Profiles With Confidence Using Conformal Prediction. *Front. Pharmacol.* **9**, 1256 (2018).
44. Eklund, M., Norinder, U., Boyer, S. & Carlsson, L. The application of conformal prediction to the drug discovery process. *Ann. Math. Artif. Intell.* **74**, 117–132 (2015).
45. Wahlberg, E. *et al.* Family-wide chemical profiling and structural analysis of PARP and tankyrase inhibitors. *Nat. Biotechnol.* **30**, 283–288 (2012).
46. Cortés-Ciriano, I., Bender, A. & Malliavin, T. Prediction of PARP Inhibition with Proteochemometric Modelling and Conformal Prediction. *Mol. Inform.* **34**, 357–366 (2015).
47. Fjodorova, N., Vračko, M., Novič, M., Roncaglioni, A. & Benfenati, E. New public QSAR model for carcinogenicity. *Chem. Cent. J.* **4**, S3 (2010).
48. Ferrari, T. & Gini, G. An open source multistep model to predict mutagenicity from statistical analysis and relevant structural alerts. *Chem. Cent. J.* **4**, S2 (2010).
49. Ahlberg, E., Spjuth, O., Hasselgren, C. & Carlsson, L. Interpretation of Conformal Prediction Classification Models. in 323–334 (Springer, Cham, 2015). doi:10.1007/978-3-319-17091-6_27
50. Cortés-Ciriano, I. *et al.* Improved large-scale prediction of growth inhibition patterns using the NCI60 cancer cell line panel. *Bioinformatics* **32**, 85–95 (2016).
51. Cortes-Ciriano, I. *et al.* Cancer Cell Line Profiler (CCLP): a webserver for the prediction of compound activity across the NCI60 panel. *bioRxiv* 105478 (2017). doi:10.1101/105478
52. Kuiper, G. G. J. M. *et al.* Comparison of the Ligand Binding Specificity and Transcript Tissue Distribution of Estrogen Receptors α and β. *Endocrinology* **138**, 863–870 (1997).
53. Taha, M. O., Tarairah, M., Zalloum, H. & Abu-Sheikha, G. Pharmacophore and QSAR modeling of estrogen receptor β ligands and subsequent validation and in silico search for new hits. *J. Mol. Graph. Model.* **28**, 383–400 (2010).
54. Norinder, U., Rybacka, A. & Andersson, P. L. Conformal prediction to define applicability domain – A case study on predicting ER and AR binding. *SAR QSAR Environ. Res.* **27**, 303–316 (2016).
55. Mysinger, M. M., Carchia, M., Irwin, J. J. & Shoichet, B. K. Directory of useful decoys, enhanced (DUD-E): better ligands and decoys for better benchmarking.

*J. Med. Chem.* **55**, 6582–6594 (2012).
56. Svensson, F., Norinder, U. & Bender, A. Improving Screening Efficiency through Iterative Screening Using Docking and Conformal Prediction. *J. Chem. Inf. Model.* **57**, 439–444 (2017).
57. Sun, J. *et al.* Applying Mondrian Cross-Conformal Prediction To Estimate Prediction Confidence on Large Imbalanced Bioactivity Data Sets. *J. Chem. Inf. Model.* **57**, 1591–1598 (2017).
58. Hansen, K. *et al.* Benchmark Data Set for in Silico Prediction of Ames Mutagenicity. *J. Chem. Inf. Model.* **49**, 2077–2081 (2009).
59. Norinder, U. & Boyer, S. Binary classification of imbalanced datasets using conformal prediction. *J. Mol. Graph. Model.* **72**, 256–265 (2017).
60. Baba, H., Takahara, J. & Mamitsuka, H. In Silico Predictions of Human Skin Permeability using Nonlinear Quantitative Structure–Property Relationship Models. *Pharm. Res.* **32**, 2360–2371 (2015).
61. Lindh, M., Karlén, A. & Norinder, U. Predicting the Rate of Skin Penetration Using an Aggregated Conformal Prediction Framework. *Mol. Pharm.* **14**, 1571–1576 (2017).
62. Svensson, F., Norinder, U. & Bender, A. Modelling compound cytotoxicity using conformal prediction and PubChem HTS data. *Toxicol. Res. (Camb).* **6**, 73–80 (2017).
63. Cortés-Ciriano, I., Bender, A., Cortes-Ciriano, I. & Bender, A. Deep Confidence: A Computationally Efficient Framework for Calculating Reliable Prediction Errors for Deep Neural Networks. *J. Chem. Inf. Model.* **59**, 1269–1281 (2019).
64. Huang, G. *et al.* Snapshot Ensembles: Train 1, get M for free. *2017, arXiv1704.00109 arXiv.org ePrint Arch. https//arxiv.org/abs/1704.00109 (accessed Jul 10, 2018).*
65. Norinder, U. *et al.* Predicting Aromatic Amine Mutagenicity with Confidence: A Case Study Using Conformal Prediction. *Biomolecules* **8**, 85 (2018).
66. Honma, M. *et al.* Improvement of quantitative structure–activity relationship (QSAR) tools for predicting Ames mutagenicity: outcomes of the Ames/QSAR International Challenge Project. *Mutagenesis* **34**, 3–16 (2019).
67. Norinder, U., Ahlberg, E. & Carlsson, L. Predicting Ames Mutagenicity Using Conformal Prediction in the Ames/QSAR International Challenge Project. *Mutagenesis* (2018). doi:10.1093/mutage/gey038
68. Norinder, U., Mucs, D., Pipping, T. & Forsby, A. Creating an efficient screening model for TRPV1 agonists using conformal prediction. *Comput. Toxicol.* **6**, 9–15 (2018).
69. Giblin, K. A., Hughes, S. J., Boyd, H., Hansson, P. & Bender, A. Prospectively Validated Proteochemometric Models for the Prediction of Small-Molecule Binding to Bromodomain Proteins. *J. Chem. Inf. Model.* **58**, 1870–1888 (2018).
70. Svensson, F., Afzal, A. M., Norinder, U. & Bender, A. Maximizing gain in high-throughput screening using conformal prediction. *J. Cheminform.* **10**, 7 (2018).
71. Johansson, H., Lindstedt, M., Albrekt, A.-S. & Borrebaeck, C. A. A genomic biomarker signature can predict skin sensitizers using a cell-based in vitro alternative to animal tests. *BMC Genomics* **12**, 399 (2011).
72. Forreryd, A., Norinder, U., Lindberg, T. & Lindstedt, M. Predicting skin sensitizers

with confidence — Using conformal prediction to determine applicability domain of GARD. *Toxicol. Vitr.* **48**, 179–187 (2018).
73. Ji, C., Svensson, F., Zoufir, A. & Bender, A. eMolTox: prediction of molecular toxicity with confidence. *Bioinformatics* **34**, 2508–2509 (2018).
74. Lapins, M. *et al.* A confidence predictor for logD using conformal regression and a support-vector machine. *J. Cheminform.* **10**, 17 (2018).
75. Dua, Dheeru and Graff, C. UCI Machine Learning Repository. *University of California, Irvine, School of Information and Computer Sciences* (2017). Available at: http://archive.ics.uci.edu/ml. (Accessed: 1st July 2019)
76. Spjuth, O., Carlsson, L. & Gauraha, N. Aggregating Predictions on Multiple Non-disclosed Datasets using Conformal Prediction. (2018).
77. Gauraha, N., Carlsson, L. & Spjuth, O. Conformal Prediction in Learning Under Privileged Information Paradigm with Applications in Drug Discovery. (2018).
78. Bosc, N. *et al.* Large scale comparison of QSAR and conformal prediction methods and their applications in drug discovery. *J. Cheminform.* **11**, 4 (2019).
79. de la Vega de León, A., Chen, B. & Gillet, V. J. Effect of missing data on multitask prediction methods. *J. Cheminform.* **10**, 26 (2018).
80. Norinder, U. & Svensson, F. Multitask Modeling with Confidence Using Matrix Factorization and Conformal Prediction. *J. Chem. Inf. Model.* acs.jcim.9b00027 (2019). doi:10.1021/acs.jcim.9b00027
81. Cortés-Ciriano, I. & Bender, A. Reliable Prediction Errors for Deep Neural Networks Using Test-Time Dropout. *J. Chem. Inf. Model.* **59**, 3330–3339 (2019).
82. Eklund, M., Norinder, U., Boyer, S. & Carlsson, L. Benchmarking Variable Selection in QSAR. *Mol. Inf.* **31**, 173–179 (2012).
83. Cortés-Ciriano, I. & Bender, A. KekuleScope: prediction of cancer cell line sensitivity and compound potency using convolutional neural networks trained on compound images. *J. Cheminform.* **11**, 41 (2019).
84. Vovk, V., Hoi, S. C. H. & Buntine, W. *Conditional validity of inductive conformal predictors*. *Mach Learn* **92**, 349–376 (Springer US, 2013).
85. Kramer, C., Kalliokoski, T., Gedeck, P. & Vulpetti, A. The experimental uncertainty of heterogeneous public K(i) data. *J. Med. Chem.* **55**, 5165–5173 (2012).
86. Cortés-Ciriano, I. & Bender, A. How consistent are publicly reported cytotoxicity data? Large-scale statistical analysis of the concordance of public independent cytotoxicity measurements. *ChemMedChem* **11**, 57–71 (2015).
87. Kalliokoski, T., Kramer, C., Vulpetti, A. & Gedeck, P. Comparability of mixed $IC_{50}$ data - a statistical analysis. *PLoS One* **8**, e61007 (2013).
88. Roberts, G., Myatt, G. J., Johnson, W. P., Cross, K. P. & Blower, P. E. LeadScope : Software for Exploring Large Sets of Screening Data. *J. Chem. Inf. Comput. Sci.* **40**, 1302–1314 (2000).
89. Beck, B., Breindl, A. & Clark, T. QM/NN QSPR Models with Error Estimation: Vapor Pressure and LogP. *J. Chem. Inf. Comput. Sci.* **40**, 1046–1051 (2000).
90. Papadopoulos, H., Vovk, V. & Gammerman, A. Regression conformal prediction with nearest neighbours. *J. Artif. Intell. Res.* **40**, 815–840 (2011).
91. Kalliokoski, T., Kramer, C. & Vulpetti, A. Quality Issues with Public Domain Chemogenomics Data. *Mol. Inform.* **32**, 898–905 (2013).
92. Cortés-Ciriano, I. *et al.* Improved large-scale prediction of growth inhibition

patterns on the NCI60 cancer cell-line panel. *Bioinformatics* **32**, 85–95 (2016).
93. Wainer, H., Gessaroli, M. & Verdi, M. Visual Revelations. *CHANCE* **19**, 49–52 (2006).
94. Sheridan, R. P. Three useful dimensions for domain applicability in QSAR models using random forest. *J. Chem. Inf. Model.* **52**, 814–823 (2012).
95. Papadopoulos, H. & Haralambous, H. Reliable prediction intervals with regression neural networks. *Neural Networks* **24**, 842–851 (2011).
96. Vovk, V. Cross-conformal predictors. *Ann. Math. Artif. Intell.* **74**, 9–28 (2015).
97. Gauraha, N. & Spjuth, O. Synergy Conformal Prediction. (2018).
98. Carlsson, L., Eklund, M. & Norinder, U. Aggregated Conformal Prediction. in 231–240 (Springer, Berlin, Heidelberg, 2014). doi:10.1007/978-3-662-44722-2_25
99. Linusson, H. *et al.* On the Calibration of Aggregated Conformal Predictors. *Proc. Mach. Learn. Res.* **60**, 1–20 (2017).
100. Ha, T.-H. *et al.* TRPV1 antagonist with high analgesic efficacy: 2-Thio pyridine C-region analogues of 2-(3-fluoro-4-methylsulfonylaminophenyl)propanamides. *Bioorg. Med. Chem.* **21**, 6657–6664 (2013).
101. Ahmed, L. *et al.* Efficient iterative virtual screening with Apache Spark and conformal prediction. *J. Cheminform.* **10**, 8 (2018).
102. Cortes-Ciriano, I. *et al.* Polypharmacology Modelling Using Proteochemometrics: Recent Developments and Future Prospects. *Med. Chem. Comm.* **6**, 24 (2015).
103. Carpenter, K. A., Cohen, D. S., Jarrell, J. T. & Huang, X. Deep learning and virtual drug screening. *Future Med. Chem.* **10**, 2557–2567 (2018).
104. Chen, H., Engkvist, O., Wang, Y., Olivecrona, M. & Blaschke, T. The rise of deep learning in drug discovery. *Drug Discov. Today* **23**, 1241–1250 (2018).
105. Niculescu-Mizil, A. & Caruana, R. Predicting good probabilities with supervised learning. in *Proceedings of the 22nd international conference on Machine learning - ICML '05* 625–632 (ACM Press, 2005). doi:10.1145/1102351.1102430
106. Srivastava, N., Hinton, G., Krizhevsky, A. & Salakhutdinov, R. Dropout: A Simple Way to Prevent Neural Networks from Overfitting. *J. Mach. Learn. Res.* **15**, 1929–1958 (2014).
107. Nowotka, M., Papadatos, G., Davies, M., Dedman, N. & Hersey, A. Want Drugs? Use Python. *2016, arXiv1607.00378 arXiv.org ePrint Arch. https//arxiv.org/abs/1607.00378 (accessed Jul 10, 2018).*
108. Mente, S. & Kuhn, M. The use of the R language for medicinal chemistry applications. *Curr. Top. Med. Chem.* **12**, 1957–1964 (2012).
109. Kuhn, M. Building Predictive Models in R Using the caret Package. *J. Stat. Softw.* **28**, 1–26 (2008).
110. Murrell, D. S. *et al.* Chemically Aware Model Builder (camb): an R package for property and bioactivity modelling of small molecules. *J. Cheminform.* **7**, 45 (2015).
111. Pedregosa, F. *et al.* Scikit-learn: Machine Learning in Python. *J. Mach. Learn. Res.* **12**, 2825–2830 (2011).
112. Paszke, A. *et al.* Automatic differentiation in PyTorch. in *Advances in Neural Information Processing Systems 30* 1–4 (2017).
113. Arvidsson, S. CPSign Documentation — CPSign 0.7.8 documentation. (2016).
114. Walters, W. P. Modeling, informatics, and the quest for reproducibility. *J. Chem.*

*Inf. Model.* **53**, 1529–1530 (2013).
115. Landrum, G. A. & Stiefl, N. Is that a scientific publication or an advertisement? Reproducibility, source code and data in the computational chemistry literature. *Future Med. Chem.* **4**, 1885–1887 (2012).
116. *Normalized Nonconformity Measures for Regression Conformal Prediction*. (Proceedings of the IASTED international conference on artificial intelligence and applications, AIA. ACTA Press, 2008).
117. Toccaceli, P. & Gammerman, A. Combination of inductive mondrian conformal predictors. *Mach. Learn.* **108**, 489–510 (2019).